\documentclass[preprint,12pt]{elsarticle}

\usepackage{slashed}
\usepackage{amssymb}
\usepackage{amsthm}
\usepackage[mathscr]{eucal}

\usepackage{color}

\journal{Nucl. Phys. B}

\begin{document}

\begin{frontmatter}

\title{Novel Topological Effects in Dense QCD in a Magnetic Field}
\author{E. J. Ferrer and V. de la Incera}

\address{Dept. of Eng. Sci. and Physics, CUNY-College of Staten Island and CUNY-Graduate Center, 
           New York 10314, USA}

\begin{abstract}
We study the electromagnetic properties of dense QCD in the so-called Magnetic Dual Chiral Density Wave phase. This inhomogeneous phase exhibits a nontrivial topology that comes from the fermion sector due to the asymmetry of the lowest Landau level modes. The nontrivial topology manifests in the electromagnetic effective action via a chiral anomaly term $~\theta F^{\mu\nu}\tilde{F}_{\mu\nu}$, with a dynamic axion field $\theta$ given by the phase of the Dual Chiral Density Wave condensate. The coupling of the axion with the electromagnetic field leads to several macroscopic effects that include, among others, an anomalous, nondissipative Hall current, an anomalous electric charge, magnetoelectricity, and the formation of a hybridized propagating mode known as an axion polariton. Connection to topological insulators and Weyls semimetals, as well as possible implications for heavy-ion collisions and neutron stars are all highlighted.
\end{abstract}


\end{frontmatter}

\section{Introduction}

Mapping the QCD phases is a goal intensely sought after by many theoretical and experimental efforts \cite{QCDreviews}. Thanks to the asymptotic freedom of the theory the most extreme regions of the QCD phase diagram in the temperature-density plane are weakly coupled and hence better understood;  they are the quark-gluon plasma (QGP) in the high-temperature/low-density corner of the phase map and the Color-Flavor-Locked (CFL) superconducting phase on the opposite site.  At low temperatures and densities quarks are confined inside hadrons, whose interactions can be phenomenologically described by conventional nuclear physics.

However, somewhere in the region of intermediate temperatures and densities, one expects a phase transition to occur from a confined to a deconfined phase, where gluons and quarks liberate from hadrons. On the other hand, confined quarks have a large dynamical mass due to the chiral condensate, while in the regions with small coupling the condensate is expected to melt and the quark mass becomes small. Clearly, by increasing the temperature or the chemical potential, a chiral phase transition should occur from the phase of heavy quarks to that of light quarks. Nevertheless, it is not obvious a priori that the confinement and chiral phase transitions have to occur at the same phase boundary line \cite{Aoki06}. 

Understanding the fundamental physics involved and the phases that realize in the intermediate density region of the QCD phase map is highly nontrivial, as it requires nonperturbative methods and effective theories. To begin with, one cannot use lattice QCD because of the sign problem at finite chemical potential, then the investigation of this region has been done with the help of NJL-like effective models \cite{NJL},  or considering QCD in the large-N limit \cite{largeNQCD}. To pin-down the correct physical description of this region, the theoretical results will have to be eventually contrasted with the outcomes of future experiments at various international facilities that plan to explore the intermediate-baryon-density and small-temperature region.

The low-temperature, intermediate-density region is particularly challenging due to the possibility of spatially inhomogeneous phases. Coming from the low density region, the energy separation between quarks and antiquarks grows with increasing density up to a point where it is not anymore energetically favorable to excite antiquarks all the way from the Dirac sea to pair them with the quarks at the Fermi surface.  Instead of undergoing a transition to a chirally restored phase, it is energetically favorable to pair quarks and holes with parallel momenta close to the Fermi surface, giving rise to inhomogeneous chiral condensates. Spatially inhomogeneous phases with quark-hole condensates have been found in the large-N limit of QCD \cite{largeNQCD}, in quarkyonic matter \cite{q-chiralspirals}, and in NJL models \cite{NickelPRD80}-\cite{PRD85-074002}. Hence, although most NJL models had predicted a first-order chiral transition with increasing density \cite{Klevansky92}, it turned out that the transition is more likely to occur via some intermediate state(s) characterized by inhomogeneous chiral condensates.

Inhomogeneous phases become favored also in color superconductivity (CS) \cite{CS-MRP80,CS-MRP86}, when the intermediate density region  is approached from the other side of the QCD phase map, that is, from the region of low temperatures and asymptotically high density values.  At asymptotic densities, the most favored CS phase is the CFL, an homogeneous phase on which all flavors pair with each other via the strong attractive quark-quark channel. Homogeneous CS is based on BCS-pairing and relies on the assumption that the quarks that pair with equal and opposite momenta can each be arbitrarily close to their common Fermi surface. However, with decreasing density, the combined effect of the strange quark mass, neutrality constraint and beta equilibrium, tends to pull apart the Fermi momenta of different flavors, imposing an extra energy cost on the formation of Cooper pairs.  BCS-pairing then dominates as long as the energy cost of forcing all species to have the same Fermi momentum is compensated by the pairing energy that is released by the formation of Cooper pairs.

Eventually, with decreasing density, homogeneous CS phases like the CFL and the 2SC become gapless and, most importantly, become unstable \cite{gCFL,g2SC}. The instability, known as chromomagnetic instability, manifests itself in the form of imaginary Meissner masses for some of the gluons and indicates an instability towards spontaneous breaking of translational invariance \cite{TBinCS}. In other words, it indicates the formation of an spatially inhomogeneous phase. Most inhomogeneous CS phases are based on the idea of Larkin and Ovchinnikov (LO) \cite{LO} and Fulde and Ferrell (FF) \cite{FF}, originally applied to condensed matter. In  the CS LOFF phases \cite{CSLOFF}, quarks of different flavors pair even though they have different Fermi momenta, because they form Cooper pairs with nonzero momentum. CS inhomogeneous phases with gluon condensates that break rotational symmetry \cite{gluonCS} have also been considered to solve the chromomagnetic instability. However, to the best of our knowledge, the question of which CS phase is the most favorable in the region of intermediate densities still remains unanswered.

The above discussion suggests that inhomogeneous phases in the low-temperature/intermediate-density region of the QCD phase map might be unavoidable, whether the region is approached from low or high densities. In this paper we are interested in the properties of a phase with a quark-hole inhomogeneous condensate that can be generated at intermediate densities in the presence of a magnetic field. Such a phase is particularly interesting because it has nontrivial topology that favors the spatial modulation of the condensate and leads to anomalous (topological) electric charge and current \cite{Ferrer-Incera-PLB}. From the onset, we admit that we cannot claim that this phase is the \textit{most} energetically favorable in the region. To answer that question requires an energy comparison of \textit{all} the possible chiral and CS phases, an insurmountable task out of the scope of this work. What we hope instead is to uncover some unique properties of this phase that can then be connected with observable signatures to establish or discard its actual realization in QCD. Because the conditions for the realization of inhomogeneous phases could be either reached at the core of compact objects or artificially produced in heavy-ion collision (HIC) experiments, any probe of these phases will have to occur in these contexts.

So far, the HIC experiments have not yet reached the QCD region of low-temperature and intermediate-densities. However, a new set of experiments planned across the globe, such as the second phase of the RHIC energy scan (BES-II) at Brookhaven in US \cite{Odyniec}, the Facility for Antiproton and Ion Research (FAIR) \cite{1607.01487} at the GSI site in Germany, and the Nuclotron-based Ion Collider Facility (NICA) \cite{PRC85} at JINR laboratory in Dubna, Russia, are aimed to extend the probe into that region in the near future. 

On the other hand, in the situations where inhomogeneous phases could be generated, magnetic fields are always present. Neutron stars typically have strong magnetic fields, which become extremely large in the case of magnetars, with inner values that have been estimated to range from $10^{18}$ G for nuclear matter \cite{Nuclear-matter-field} to $10^{20}$ G for quark matter \cite{Quark-Matter-field}. Likewise, off-central HIC are known to produce large magnetic fields. Their values have been estimated to be of order eB $\simeq O(2m^2_\pi) \sim10^{18}$G for the top collision in non-central Au-Au impacts at RHIC, and even larger, eB $\simeq$ $O(15m^2_\pi) \sim 10^{19}$G, at LHC \cite{CME, Skokov}. Even though the magnetic field produced in HIC is transient and inhomogeneous; within the characteristic length scale and life time of the QGP formed in HIC, it is reliable to consider the magnetic field as approximately constant and homogeneous (see discussion in \cite{Wen} and references therein).

The presence of a magnetic field is relevant because it can significantly enhance the window for inhomogeneous phases \cite{KlimenkoPRD82, PLB743}. Moreover, a magnetic field can activate attractive channels and produce new condensates as a consequence of the explicit breaking of the rotational symmetry, as occurs, for instance, in the case of a constant chiral condensate \cite{AMM}, in color superconductivity \cite{ferrer-incera, Lec-Notes} and in quarkyonic matter \cite{ferrer-incera-sanchez}.  

In this paper, we study the electromagnetic properties of the so-called Magnetic Dual Chiral Density Wave (MDCDW) phase \cite{Ferrer-Incera-PLB} of quarks at high density in the presence of a magnetic field. This phase breaks chiral and translational symmetry and is characterized by  asymmetric fermion energy modes in the lowest Landau level (LLL). We present detailed calculations of the results discussed in a recent letter \cite{Ferrer-Incera-PLB}, as well as new derivations that provide deeper insight in the physics of the MDCDW phase. We also discuss the magnetoelectricity of this phase and how the anomalous Hall current can be found from the effective magnetic current $\nabla \times \mathbf{M}$.

Despite marked differences between strongly interacting quarks at high density and condensed matter materials, we have uncovered striking similarities in the transport properties of the MDCDW phase and those of topological materials like Weyl semimetals (WSM) \cite{Weyl-SM}, which are conductive materials with a nontrivial electronic structure topology. This is a particularly exciting finding, in light of the recent booming of works dedicated to the topic of topological materials, and the realization that they are not limited to two-dimensional systems \cite{burkov-2017}. Thanks to their similarity, we can use tabletop experiments with WSM to mimic effects that could happen in the MDCDW phase but would be much harder to detect in a HIC experiment or in astrophysical observations. Equally, we could take advantage of new understandings within these materials to infer potentially measurable effects in the MDCDW phase of quark matter, and then use that insight to design clever ways to probe the presence of this quark phase in neutron stars and in HIC experiments.

Although the origin of the nontrivial topology in both, the MDCDW phase and the WSM, comes from the fermion sector, they have very different nature. In the MDCDW the topology is connected to the inhomogeneous chiral condensate in the magnetic field, while in the WSM it comes from the existence of Weyl nodes with opposite chiralities and separated in momentum space. Nevertheless, the two systems share a basic common element, the presence of the chiral anomaly  $\frac{\kappa}{4}\theta F_{\mu\nu}\widetilde{F}^{\mu\nu}$ (axion term) in the effective action. The chiral anomaly couples the axion field $\theta$ with the electric and magnetic fields. It appears because of the lack of invariance of the fermion measure under a chiral gauge symmetry transformation. In the quark case, $\nabla\theta$ is associated with the condensate modulation, while in the WSM, it is related to the momentum separation of the Weyl nodes. In both systems, the electromagnetism is characterized by the equations of Axion Electrodynamics, so they exhibit similar electric transport properties \cite{Ferrer-Incera-PLB}.
 
We highlight that in the MDCDW phase, the interplay between the magnetic field and the inhomogeneous condensate is essential to produce the anomalous properties of the system. Without these two elements combined, there will be no LLL spectral asymmetry, a property that is key to generate the non-trivial topology that is subsequently reflected in the presence of the chiral anomaly.

The paper is organized as follows. In Section 2, the NJL model that leads to the MDCDW condensate is reexamined. In Section 3, we take into account that the fermion measure in the path integral is not invariant under a chiral gauge transformation and use the Fujikawa's method to regularize it in a gauge invariant way and to find its finite contribution to the electromagnetic effective action. The equations of axion electrodynamics are derived and the anomalous terms discussed in Section 4. In Section 5, we explore the topological origin of the anomalous electric charge and current. Section 6 is dedicated to calculate all the ordinary charges and current densities in the MDCDW phase and to demonstrate that they do not cancel the anomalous contributions. The magnetoelectric  properties of the MDCDW phase are discussed in Section 7; while Section 8 analyzes the anomalous transport effects of this phase. In Section 9, we give our concluding remarks.

\section{ Model of Cold and Dense Quark Matter in a Magnetic Field}

 Henceforth, we focus on the cold and dense region of QCD. A growing body of works indicates that with increasing density, the chirally broken phase of quark matter is not necessarily replaced by a chirally restored phase, but instead, at least for the region of intermediate densities, the system may favor the formation of spatially inhomogeneous phases. To understand this, notice that with increasing density the homogeneous chiral condensate becomes disfavored due to the high-energy cost of exciting the antiquarks from the Dirac sea to the Fermi surface where the pairs form. At the same time, with higher density, co-moving quarks and holes at the Fermi surface may pair with minimal energy cost through a mechanism analogous to Overhauser's \cite{Overhauser}, thereby giving rise to a spatially modulated condensate \cite{largeNQCD}. Spatially modulated chiral condensates have been discussed in QCD within the context of quarkyonic matter \cite{QqM}, where they appear in the form of quarkyonic chiral spirals \cite{q-chiralspirals} at zero magnetic field, or double quarkyonic chiral spirals  \cite{ferrer-incera-sanchez} in the presence of a magnetic field. Inhomogeneous chiral condensates have been also studied in NJL models (for a review see \cite{InhCRev}) that share the chiral symmetries of QCD and are then useful to investigate the chiral phase transition. 
 
We are interested in the electromagnetic properties of cold and dense quark matter in a background magnetic field. With that goal in mind, we model QCD+QED with the help of the following Lagrangian density that combines electromagnetism with a two-flavor NJL model of strongly interacting quarks,
\begin{eqnarray} \label{L_NJL_QED}
\mathcal{L}=&-&\frac{1}{4}F_{\mu\nu}F^{\mu\nu}+\bar{\psi}[i\gamma^{\mu}(\partial_\mu+iQA_{\mu})+\gamma_0 \mu]\psi 
\nonumber
\\
&+&G[(\bar{\psi}\psi)^2+(\bar{\psi}i\mathbf{\tau}\gamma_5\psi)^2],
\end{eqnarray}
Here $Q=\mathrm{diag} (e_u,e_d)=\mathrm{diag} (\frac{2}{3}e,-\frac{1}{3}e)$, $\psi^T=(u,d)$; $\mu$ is the baryon chemical potential; and G is the four-fermion coupling. The electromagnetic potential $A^{\mu}$ is formed by the background $\bar{A}^{\mu}=(0,0,Bx,0)$, which corresponds to a constant and uniform magnetic field $\mathbf{B}$ pointing in the z-direction, with $x^{\mu}=(t,x,y,z)$, plus the fluctuation field $\tilde{A}$. We work in the metric $g_{\mu\nu}=\mathrm{diag}(1,-\mathbf{1})$. Due to the electromagnetic coupling of the quarks, the flavor symmetry $SU(2)_L\times SU(2)_R$ of the original NJL model is reduced to the subgroup $U(1)_L\times U(1)_R $.

The presence of $\mathbf{B}$ favors the formation of a dual chiral density wave (DCDW) condensate  \begin{equation}
\langle\bar{\psi}\psi\rangle= \Delta \cos q_{\mu}x^{\mu}, \qquad \langle\bar{\psi}i\tau_3 \gamma_5\psi\rangle= \Delta \sin{q_{\mu}x^{\mu}} 
\end{equation}
with magnitude $\Delta$ and modulation $q^{\mu}=(0,0,0,q)$ along the field direction \cite{KlimenkoPRD82,PLB743}. As it will become apparent throughout the paper, the two-flavor system with the DCDW condensate in the presence of a magnetic field  \cite{KlimenkoPRD82,PLB743,Ferrer-Incera-PLB} not only differs in its symmetry from the DCDW phase at zero magnetic field  \cite{DCDW}, but it also exhibits very different transport properties. Hence the special name of MDCDW \cite{Ferrer-Incera-PLB} for this new phase.

The mean-field Lagrangian of the MDCDW phase is
\begin{eqnarray} \label{DCDW-MF_L}
\mathcal{L}_{MF}&=&\bar{\psi}[i\gamma^{\mu}(\partial_\mu+iQA_{\mu})+\gamma_0\mu]\psi-m\bar{\psi}e^{i\tau_3\gamma_5q_{\mu}x^{\mu}}\psi
\nonumber
\\
&-&\frac{m^2}{4G}-\frac{1}{4}F_{\mu\nu}F^{\mu\nu},
\end{eqnarray}
where $m=-2G\Delta$. 

To remove the spatial modulation of the mass, we use a local chiral transformation
\begin{equation}\label{Chiral-Transformations}
\psi \to U_A\psi, \quad \bar{\psi} \to \bar{\psi}\bar{U}_A, 
\end{equation}
with $U_A=e^{i\tau_3\gamma_5\theta}$, $\bar{U}_A=\gamma_0U^\dag\gamma_0=e^{i\tau_3\gamma_5 \theta}$, and $\theta(x)=qz/2$.

After the chiral transformation (\ref{Chiral-Transformations}), the mean-field Lagrangian density (\ref{DCDW-MF_L}) becomes 
\begin{eqnarray}\label{U_1-MF_L}
\mathcal{L}_{MF}&=&\bar{\psi}[i\gamma^{\mu}(\partial_\mu+iQA_{\mu}+i\tau_3\gamma_5\partial_{\mu}\theta) +\gamma_0\mu-m]\psi 
\nonumber
\\
&-&\frac{m^2}{4G}-\frac{1}{4}F_{\mu\nu}F^{\mu\nu}
\end{eqnarray}

Accordingly, the fermion contribution to the mean-field action is given by  \cite{KlimenkoPRD82}
\begin{equation}\label{mf-Z}
\Gamma=-i\log Z=\sum_f \frac{N_c}{i} \ln \mathrm{Det}(i\partial_0+\mu-H_f)-\int d^4x (\frac{m^2}{4G})
\end{equation}
where the quark partition function is
\begin{equation}\label{Z}
Z= \int \mathcal{D}\bar{\psi}(x)\mathcal{D}\psi(x) e^{i\int d^4x\lbrace\bar{\psi}[i\gamma^{\mu}(\partial_\mu+iQA_{\mu}+i\tau_3\gamma_5\partial_{\mu}\theta) +\gamma_0\mu-m]\psi-\frac{m^2}{4G}\rbrace}
\end{equation}
and
\begin{equation}\label{Hf} 
H_f=-i\gamma^0\gamma^{i}(\partial_i+ie_fA_{i}+i\frac{e_f}{|e_f|}\gamma_5\partial_{i}\theta)+\gamma^0m,
\end{equation}
denotes the modified Dirac Hamiltonian of flavor $f$ in the MDCDW phase. Here $e_f$ is the flavor electric charge.

The single-particle energy spectrum is given by the eigenvalues of $H_f$ \cite{KlimenkoPRD82}. It separates into two set of energy modes, the LLL ($l=0$) modes
\begin{equation}\label{LLLspectrum}
E^{0}=\epsilon\sqrt{m^2+k_3^2}+q/2,  \quad \epsilon=\pm,
\end{equation}
and the higher Landau level ($l\neq0$) modes
\begin{equation}\label{HighLspectrum}
E^l= \epsilon\sqrt{(\xi\sqrt{m^2+k_3^2}+q/2)^2+2|e_fB|l}, \quad \epsilon=\pm, \xi=\pm, l=1,2,3,...
\end{equation}
In (\ref{HighLspectrum}) $\xi=\pm$ indicates spin projection and $\epsilon=\pm$ particle/antiparticle energies. In contrast, only one spin projection ($+$ for positively charged and $-$ for negatively charged quarks) contributes to the LLL spectrum. An important feature of this spectrum is that the LLL energies are not symmetric about the zero-energy level. As a consequence, the $\pm$ sign in front of the square root should not be interpreted as particle/antiparticle in the LLL case.

\section{Non-invariance of the Measure and Modification of the Effective Action}

In this section, we turn our attention to a very important fact: the fermion measure in the path-integral is not invariant under the chiral gauge transformation (\ref{Chiral-Transformations}), 
\begin{equation}\label{MeasureNI}
\mathcal{D}\bar{\psi}(x)\mathcal{D}\psi(x) \to J_{\bar{\psi}} J_{\psi}\mathcal{D}\bar{\psi}(x)\mathcal{D}\psi(x).
\end{equation}
To take this into consideration we need to calculate the Jacobian $J_{\psi}=J_{\bar{\psi}}=(\textrm{Det}U_A)^{-1}$.

Using the relation $\langle x|\mathcal{O} |y\rangle=\delta^4(x-y) \mathcal{O}(x)$, valid for ultra-local integral kernels, we can show that
\begin{eqnarray}\label{Ill-defDet}
( \textrm{Det}U_A)^{-1}&=&e^{-\mathrm{Tr} \ln U_A}=e^{-\int d^4x \langle x |\mathrm{tr} \ln U_A|x\rangle} \nonumber
\\
&=&e^{-N_c\int d^4x \delta^4(0) i\theta(x) \mathrm{tr} (\gamma_5\tau_3)},
\end{eqnarray}
where $\mathrm{Tr}$ indicates functional+matrix trace, while $\mathrm{tr}$ indicates just matrix trace. It is easy to see that the exponent in (\ref{Ill-defDet}) is an ill-defined conditionally convergent quantity that needs to be properly regularized.

With that goal in mind, we follow the Fujikawa's approach \cite{FujikawaPRD21_1980}, on which the Jacobian is regularized in a gauge-invariant way as described in details below. The essence of the method consists of first expressing the Jacobian in the representation of the eigenfunctions of an Euclidean operator that is gauge-invariant and Hermitian (antiHermitian).  Such a representation not only preserves the gauge invariance of the theory, but also ensures that the eigenfunctions are orthogonal and complete  and have real (imaginary) eigenvalues. Additionally, the functional space must be chosen so that it diagonalizes the fermion action. Such a diagonalization is essential to ensure the unitarity of the transformation from the original fermion fields to the eigenvectors of the chosen operator in the presence of regularization. Notice that because of the ill-defined Jacobian, a seemingly unitary transformation based on the eigenspace of a gauge invariant operator that does not diagonalize the action is actually nonunitary \cite{FujikawaPRD21_1980}-\cite{JHEP1112_2011}. Further, to regularize the Jacobian one introduces a damping factor in the form of an arbitrary function of the eigenvalues of this operator with a regulator $M$, in such a way that the contributions from the large momenta are regularized when $M \to \infty$. Below, we use the heat-kernel regularization \cite{Nakahara}, which is based on an exponential damping function.

In most cases, the gauge-invariant operator whose eigenfunctions satisfy all the above requirements is the Dirac operator of the theory $\slashed{D}$ \cite{Bilal}. However, in our case, the presence of the chemical potential and the chiral term $\sim \gamma_5 \partial_\mu \theta$ in the covariant derivative spoil the Hermiticity of the Dirac operator in the Euclidean space and the Fujikawa approach has to be extended.  In preparation for this process we first perform a Wick rotation to Euclidean space  $dx_0 \rightarrow -idx_4$, $\partial_0 \rightarrow i\partial_4$, $A_0\rightarrow iA_4$, and for the Dirac matrices, $\gamma_0\rightarrow i\gamma_4$, $\gamma^5 \equiv i\gamma^0 \gamma^1 \gamma^2 \gamma^3=\gamma_E^5=\gamma^1 \gamma^2 \gamma^3 \gamma^4$. The Euclidean $\gamma_{\mu}$ are all anti-Hermitian and the Euclidean metric becomes in this case $g^E_{\mu\nu}= \mathrm{diag}(-1,-1,-1,-1)$. 

Notice that, since there is no mix between the quark flavors, we can perform, without loss of generality, the analysis for one quark flavor and then incorporate the contributions of the two flavors in the final result. In the present system, the Euclidean Dirac operator is $\slashed{D}(\mu,\theta)=\slashed{D}+\slashed{D}^A$, with $\slashed{D}=\gamma_\mu(\partial_\mu+ie_fA_\mu)$ Hermitian, and $\slashed{D}^A=\gamma_\mu(i \gamma^5\mathrm{sgn}(e_f)\partial_{\mu}\theta-\mu \delta_{\mu4})$ anti-Hermitian. Since $\slashed{D}(\mu,\theta)$ is neither Hermitian nor anti-Hermitian, it cannot be used to apply the Fujikawa approach. In this case, we follow instead the method discussed in \cite{JHEP1112_2011}, which extended Fujikawa's approach to Dirac operators at finite density. Because of the chiral term, we have to extend that method even further and consider not just the chemical potential but also  the axial "gauge field" $\partial_\mu \theta$ in the covariant derivative.  

Consider the positive-semidefinite Hermitian operators $\slashed{D}^\dag(\mu,\theta)\slashed{D}(\mu,\theta)$ and $\slashed{D}(\mu,\theta)\slashed{D}^\dag(\mu,\theta)$ and their respective eigenvalue equations
\begin{equation}\label{DdagDeigenf}
\slashed{D}^\dag(\mu,\theta)\slashed{D}(\mu,\theta)\phi_n=\lambda_n^2\phi_n \qquad \slashed{D}(\mu,\theta)\slashed{D}^\dag(\mu,\theta)\widetilde{\phi}_n=\xi_n^2\widetilde{\phi}_n, 
\end{equation}
whose eigenfunctions form sets that are complete
\begin{equation}\label{Completeness}
\sum_n \phi^\dag_n(x)\phi_n(y)=\delta(x-y) \qquad \sum_n \widetilde{\phi}^\dag_n(x)\widetilde{\phi}_n(y)=\delta(x-y)
\end{equation}
and orthogonal
\begin{equation}\label{Orthogonality}
\int d_E^4x{\phi}^\dag_n(x)\phi_m(x)=\delta_{nm}, \qquad \int d_E^4x\widetilde{\phi}^\dag_n(x)\widetilde{\phi}_n(x)=\delta_{nm}.
\end{equation} 
and have real eigenvalues $\lambda_n$, $\xi_n$, known as the singular values of $\slashed{D}(\mu,\theta)$, $\slashed{D}^\dag(\mu,\theta)$ respectively. The $\phi_n(x)$, $\widetilde{\phi}_n(x)$ are ordinary c-number functions.
%and that $\widetilde{\phi}_n \sim \slashed{D}(\mu,\theta)\phi_n$ and $\phi_n \sim \slashed{D}^\dag(\mu,\theta)\widetilde{\phi}_n$.
%by acting with $\slashed{D}(\mu,\theta)$ on the first equation of (\ref{DdagDeigenf}) and with $\slashed{D}^\dag(\mu,\theta)$ on the second 

It is easy to verify, as in the case studied in \cite{JHEP1112_2011}, that the operators $\slashed{D}^\dag(\mu,\theta)\slashed{D}(\mu,\theta)$ and $\slashed{D}(\mu,\theta)\slashed{D}^\dag(\mu,\theta)$ share all the nonzero eigenvalues. To see this, consider a nonzero $\lambda_n$ and let us act with $\slashed{D}(\mu,\theta)$ on the first equation of (\ref{DdagDeigenf})
\begin{equation}\label{Dnuevo}
\slashed{D}(\mu,\theta)\slashed{D}^\dag(\mu,\theta)\slashed{D}(\mu,\theta)\phi_n=\lambda_n^2\slashed{D}(\mu,\theta)\phi_n
\end{equation}
That means that $\slashed{D}(\mu,\theta)\phi_n$ is an eigenfunction $\widetilde{\phi}_n$ of $\slashed{D}(\mu,\theta)\slashed{D}^\dag(\mu,\theta)$ with eigenvalue $\lambda_n^2$. Similarly, acting with $\slashed{D}^\dag(\mu,\theta)$ on the second equation of (\ref{DdagDeigenf}) we find
\begin{equation}\label{Dnuevo1} 
\slashed{D}^\dag(\mu,\theta)\slashed{D}(\mu,\theta)\slashed{D}^\dag(\mu,\theta)\widetilde{\phi}_n=\xi_n^2\slashed{D}^\dag(\mu,\theta)\widetilde{\phi}_n, 
\end{equation}
we see that $\slashed{D}^\dag(\mu,\theta)\widetilde{\phi}_n$ is an eigenfunction $\phi_n$ of $\slashed{D}^\dag(\mu,\theta)\slashed{D}(\mu,\theta)$ with eigenvalues $\xi_n^2$, so  $\lambda_n^2=\xi_n^2$.  Hence, we define from now on, for nonzero $\lambda_n$, $\widetilde{\phi}_n=\lambda^{-1}_n \slashed{D}(\mu,\theta)\phi_n$ . 

It is convenient to expand the fermion fields in the path integral in the bases of the Hermitian operators $\slashed{D}^\dag(\mu,\theta)\slashed{D}(\mu,\theta)$ and $\slashed{D}(\mu,\theta)\slashed{D}(\mu,\theta)^\dag$,

\begin{equation}\label{representation}
\psi(x)=\sum_{n} a_n\phi_n(x), \qquad \qquad \bar{\psi}(x)=\sum_{n} \bar{b}_n{\widetilde{\phi}}^\dag_n(x),
\end{equation}
with $a_n$, $b_n$, Grassmann numbers. In the representation of these eigenfunctions, the Jacobian of flavor $f$ in (\ref{MeasureNI}) takes the form

\begin{equation}\label{Jacobian-phi-rep}
J^{(f)}_{\psi}J^{(f)}_{\bar{\psi}}=e^{iN_c\mathrm{tr}\int d_E^4x \theta(x) \sum_{n} [\phi^\dag_n(x) i\gamma_5 \phi_n(x)+\widetilde{\phi}^\dag_n(x) i\gamma_5 \widetilde{\phi}_n(x)].}
\end{equation}
and the fermionic part of the action is diagonalized
\begin{equation} \label{diagonal-action}
S_F=\int d_E ^4x\bar{\psi}\slashed{D}(\mu,\theta)\psi=\sum_{n}\lambda_n \bar{b}_n a_n.
\end{equation}

We now turn to the standard heat-kernel regularization method \cite{Nakahara}, and introduce damping factors for each term in (\ref{Jacobian-phi-rep}) and a regulator $M$ that will be taken to infinity at the end. The regularized Jacobian then becomes

\begin{equation}\label{Jacobian-phi-rep_Reg}
 J^{(f)}_{\psi}J^{(f)}_{\bar{\psi}}
 =e^{iN_c(\mathcal{I}_R+\mathcal{\widetilde{I}}_R)}
\end{equation}
where 
\begin{eqnarray}\label{Heat-Kernel}
\mathcal{I}_R&=&\lim_{M \rightarrow \infty} \int d_E^4x \theta(x)\mathrm{tr}\sum_n\phi_n^{\dag}(x)  i\gamma_5 e^{-\lambda_n^2/M^2} \phi_n(x)  \nonumber
\\
&=&\lim_{M \rightarrow \infty} \int d_E^4x \theta(x)\mathrm{tr}\sum_n\phi_n^{\dag}(x) i \gamma_5 e^{-\slashed{D}^\dag(\mu,\theta)\slashed{D}(\mu,\theta)/M^2}\phi_n(x) \nonumber
\\
&\equiv& \lim_{M \rightarrow \infty}\int d_E^4x \theta(x)I,
\end{eqnarray}
and
\begin{eqnarray}\label{Heat-Kernel1}
\mathcal{\widetilde{I}}_R&=&\lim_{M \rightarrow \infty} \int d_E^4x \theta(x)\mathrm{tr}\sum_n\widetilde{\phi}_n^{\dag}(x)  i\gamma_5 e^{-\lambda_n^2/M^2} \widetilde{\phi}_n(x)  \nonumber
\\
&=&\lim_{M \rightarrow \infty} \int d_E^4x \theta(x)\mathrm{tr}\sum_n \widetilde{\phi}_n^{\dag}(x) i \gamma_5 e^{-\slashed{D}(\mu,\theta)\slashed{D}^\dag(\mu,\theta)/M^2}\widetilde{\phi}_n(x) \nonumber
\\
&\equiv& \lim_{M \rightarrow \infty}\int d_E^4x \theta(x)\widetilde{I},
\end{eqnarray}
with
\begin{eqnarray}\label{Square-Opr}
\slashed{D}^\dag(\mu,\theta)\slashed{D}(\mu,\theta)&=&-(D_\mu)^2+\frac{ie_f}{4} [\gamma_\mu, \gamma_\nu] F_{\mu\nu}+(\partial_{\mu}\theta)^2-i\mathrm{sgn}(e_f)\gamma_5[\gamma_\mu,\gamma_4]\mu\partial_\mu\theta+\mu^2 \nonumber
\\
&+&i \mathrm{sgn}(e_f)[\gamma_\mu, \gamma_\nu]\gamma_5\partial_\mu\theta D_{\nu}+[\gamma_4,\gamma_\nu]\mu D_\nu
\end{eqnarray}
\begin{eqnarray}\label{Square-Opr1}
\slashed{D}(\mu,\theta)\slashed{D}^\dag(\mu,\theta)&=&-(D_\mu)^2+\frac{ie_f}{4} [\gamma_\mu, \gamma_\nu] F_{\mu\nu}+(\partial_{\mu}\theta)^2-i\mathrm{sgn}(e_f)\gamma_5[\gamma_\mu,\gamma_4]\mu\partial_\mu\theta+\mu^2 \nonumber
\\
&-&i \mathrm{sgn}(e_f)[\gamma_\mu, \gamma_\nu]\gamma_5\partial_\mu\theta D_{\nu}-[\gamma_4,\gamma_\nu]\mu D_\nu
\end{eqnarray}
Here $D_\mu\equiv \partial_\mu+ie_f A_\mu$  and we used that $[D_\mu, D_\nu]=ie_fF_{\mu\nu}$. 

Once the Jacobian is regularized, it is convenient to change the basis to the free-wave eigenfunctions  $|\zeta \rangle$ $,\left(\slashed{\partial}|  \zeta \rangle=i\slashed{k}|  \zeta \rangle \right)$, to find

\begin{eqnarray}\label{Square-Opr-2}
I&=& \mathrm{tr}\sum_{n} \langle \phi_n |  x \rangle i \gamma_5 e^{-\slashed{D}^\dag(\mu,\theta)\slashed{D}(\mu,\theta)/M^2}  \langle x|  \phi_n \rangle  \qquad \qquad \qquad \qquad\ \nonumber
\\
&=& \mathrm{tr} \int \frac{d^4k}{(2\pi)^4}\int \frac{d^4k'}{(2\pi)^4}\sum_{n} \langle \phi_n |  \zeta \rangle   \langle \zeta |  x \rangle  i \gamma_5 e^{-\slashed{D}^\dag(\mu,\theta)\slashed{D}(\mu,\theta)/M^2}  \langle x|  \zeta' \rangle\langle \zeta'|  \phi_n \rangle \ \nonumber
\\
&=& \mathrm{tr}\int \frac{d^4k}{(2\pi)^4}\int \frac{d^4k'}{(2\pi)^4} \langle \zeta' |  \zeta \rangle   \langle \zeta |  x \rangle  i \gamma_5 e^{-\slashed{D}^\dag(\mu,\theta)\slashed{D}(\mu,\theta)/M^2}  \langle x|  \zeta' \rangle  \qquad  \qquad  \nonumber
\\
&=&\mathrm{tr}\int \frac{d^4k}{(2\pi)^4} e^{-ikx}  i \gamma_5 e^{-\slashed{D}^\dag(\mu,\theta)\slashed{D}(\mu,\theta)/M^2}  e^{ikx} \nonumber
\\ 
 &=&\mathrm{tr}\int \frac{d^4k}{(2\pi)^4} i \gamma_5 e^{-\slashed{D}^\dag(k,\mu,\theta)\slashed{D}(k,\mu,\theta)/M^2} , 
\end{eqnarray}
and similarly for $\widetilde{I}$

\begin{eqnarray}\label{Square-Opr-3}
\widetilde{I}&=& \mathrm{tr}\sum_{n} \langle \widetilde{\phi}_n |  x \rangle i \gamma_5 e^{-\slashed{D}(\mu,\theta)\slashed{D}^\dag(\mu,\theta)/M^2}  \langle x|  \widetilde{\phi}_n \rangle  \qquad \qquad \qquad \qquad\ \nonumber
\\
 &=&\mathrm{tr}\int \frac{d^4k}{(2\pi)^4} i \gamma_5 e^{-\slashed{D}(k,\mu,\theta)\slashed{D}^\dag(k,\mu,\theta)/M^2} , 
\end{eqnarray}
with $\slashed{D}^\dag(k,\mu,\theta)\slashed{D}(k,\mu,\theta)$ and $\slashed{D}(k,\mu,\theta)\slashed{D}^\dag(k,\mu,\theta)$ given respectively by (\ref{Square-Opr}) and (\ref{Square-Opr1}) with $D_\mu$ replaced by $(ik_\mu+D_\mu)$.

At this point, we make the variable change $k_\mu \rightarrow Mk_\mu$ in (\ref{Square-Opr-2})  and (\ref{Square-Opr-3}), use them back in (\ref{Heat-Kernel}) and (\ref{Heat-Kernel1}), and take the trace and the limit $M \to \infty$. Then, we can readily verify that  
\begin{eqnarray}\label{Trace}
\mathcal{I}_R+ \mathcal{\widetilde{I}}_R&=& -\frac{2e_f^2}{32}\int d_E^4x \theta(x) \int \frac{d^4k}{(2\pi)^4}e^{-k^2}\mathrm{tr}i\gamma^5[\gamma_\mu, \gamma_\nu][\gamma_\alpha, \gamma_\beta]F_{\mu\nu}F_{\alpha \beta} \nonumber
\\
&=& -\frac{e_f^2}{16}\int d_E^4x  \theta(x) \frac{1}{16\pi^2}\mathrm{tr}i\gamma^5[\gamma_\mu, \gamma_\nu][\gamma_\alpha, \gamma_\beta]F_{\mu\nu}F_{\alpha \beta}\nonumber
\\
&=&- i\frac{e_f^2}{8\pi^2}\int d^4x \theta(x)F_{\mu\nu}\tilde{F}^{\mu \nu},
\end{eqnarray}
where in the last line of (\ref{Trace}) we Wick rotated back to Minkowski space and used $\mathrm{tr}\gamma^5[\gamma^\mu, \gamma^\nu][\gamma^\alpha, \gamma^\beta]=-16i\epsilon^{\mu \nu \alpha \beta}$, with $\epsilon^{\mu \nu \alpha \beta}$ the Levi-Civita tensor and $\tilde{F}^{\mu \nu}=\frac
{1}{2}\epsilon^{\mu \nu \alpha \beta}F_{\mu\nu}$ the dual electromagnetic tensor.

Hence, the one-flavor Jacobian reduces to
\begin{equation}\label{Jacobian-2}
J^{(f)}_{\psi}J^{(f)}_{\bar{\psi}}
 =e^{iN_c(\mathcal{I}_R+\mathcal{\widetilde{I}}_R)}=e^{N_c\frac{e_f^2}{8\pi^2} \int d^4x  \theta(x) F_{\mu\nu}\tilde{F}^{\mu \nu}}
\end{equation} 
Considering the contribution of the two flavors, we finally find the regularized Jacobian of the chiral transformation to be
\begin{equation}\label{Jacobian-final}
( \det U_A)^{-2}=e^{\frac{e^2}{8\pi^2} \int d^4x  \theta(x) F_{\mu\nu}\tilde{F}^{\mu \nu}}
\end{equation} 

In Section 5, we will see the connection between (\ref{Jacobian-final}) and the chiral anomaly. The presence of the chemical potential does not affect the Jacobian or the chiral anomaly in the MDCDW phase, in agreement with similar results found in other models discussed in the literature \cite{JHEP1112_2011},\cite{anomalywithmu}.

The result (\ref{Jacobian-final}) is regularization-independent, as can be seen by replacing the exponential $e^{-\lambda^2/M^2}$ in (\ref{Heat-Kernel}) by $f((\lambda/M)^2)$, with $f(t)$ an arbitrary and smooth function, such that $f(0)=1$, $f(\infty)=0$, $tf'(t)=0$ at $t=0$ and $t=\infty$. It is straightforward to demonstrate that in this case one arrives exactly at the same result \cite{FujikawaPRD21_1980,VI-LAquila}.

Taking into account (\ref{Jacobian-final}), the quark partition function can now be written as
\begin{equation}\label{Z-theta}
Z=e^{i\Gamma_{matter}}= \int \mathcal{D}\bar{\psi}(x)\mathcal{D}\psi(x) e^{iS_{eff}}
\end{equation}
with 
 \begin{eqnarray}\label{Action-Effective}
S_{eff}&=&\int d^4x \{\bar{\psi}[i\gamma^{\mu}(\partial_\mu+iQA_{\mu}+i\tau_3\gamma_5\partial_{\mu}\theta) +\gamma_0\mu-m]\psi
-\frac{m^2}{4G}\nonumber
\\
&+&\frac{\kappa}{4} \theta(x) F_{\mu\nu}\tilde{F}^{\mu \nu}\},
\end{eqnarray}
with  $\frac{\kappa}{4}=\frac{3(e_u^2-e_d^2)}{8\pi^2}=\frac{e^2}{8\pi^2}=\frac{\alpha}{2\pi}$. Note that the axion term $\frac{\kappa}{4} \theta(x) F_{\mu\nu}\tilde{F}^{\mu \nu}$ is of order $\alpha$. Such a term, whose important physical implications will be discussed in the next sections, was overlooked by other authors who studied the MDCDW phase. 

\section{Axion Electrodynamics in the MDCDW Phase}

To find the zero temperature electromagnetic effective action $\Gamma(A)$, we first start from the one-loop effective action for the fermions\begin{equation}
\Gamma=-i\log Z,
\end{equation}
with $Z$ given by (\ref{Z-theta}), go to Euclidean variables $k^0\to ik^4$ and then employ the finite-temperature Matsubara technique
\begin{eqnarray}
&\int _{-\infty}^{\infty}&\frac{dk^0}{2\pi} \to i\int _{-\infty}^{\infty}\frac{dk_4}{2\pi} \to  i\frac{1}{\beta}\sum_{n=-\infty}^{\infty} \nonumber
\\
&k^4 &\to \omega_n=\frac{2\pi(n+\frac{1}{2})}{\beta}, \quad n=0, \pm1, \pm2, \cdots\nonumber
\end{eqnarray} 
with $\beta=1/T$ the inverse absolute temperature. We then integrate in the fermion fields, expand $\Gamma$ in powers of the fluctuation field $\tilde{A}$, sum in the Matsubara index $n$ and take the zero-temperature limit. 

At this point, we just need to add the Maxwell term to $\Gamma$ to obtain 
\begin{eqnarray} \label{EA-2}
\Gamma(A)&=&-V\Omega+\int d^4x \left[-\frac{1}{4}F_{\mu\nu}F^{\mu\nu}+\frac{\kappa}{4} \theta(x) F_{\mu\nu}\tilde{F}^{\mu \nu}\right]
\\
&+ &\sum_{i=1}^{\infty}\int dx_{1}...dx_{i}\Pi^{\mu_1,\mu_2,...\mu_i}(x_1,x_2,...x_i)\tilde{A}_{\mu_1}(x_1)...\tilde{A}_{\mu_i}(x_i),\nonumber
\end{eqnarray}
with V the four-volume, $\Omega$ the mean-field thermodynamic potential obtained for this phase in Ref. \cite{KlimenkoPRD82} (see Section 6.3 below), and $\Pi^{\mu_1,\mu_2,...\mu_i}$ the i-vertex tensors corresponding to the one-loop polarization operators with internal lines of fermion Green functions in the MDCDW phase and $i$ external lines of photons.

We are interested in the linear response of the MCDCW phase to a small electromagnetic probe $\tilde{A}$ in the background of a magnetic field $\vec{B}$ that does not need to be large either, since there is no critical magnetic field for the MDCDW condensate to form. Furthermore, for consistency of the approximation, we can neglect all the radiative corrections of order higher than $\alpha$, as $\alpha$ is the order of the axion term in (\ref{EA-2}).  These two conditions imply that we shall cut the series in (\ref{EA-2}) at $i=1$, which can be shown to provide the medium corrections to the Maxwell equations that are linear in the electromagnetic field and of the desired order in $\alpha$.

Then, $\Gamma(A)$ becomes  
\begin{eqnarray} \label{EA}
\Gamma(A)&=&-V\Omega+\int d^4x \left[-\frac{1}{4}F_{\mu\nu}F^{\mu\nu}-\kappa\int d^4x  \epsilon^{\mu \alpha \nu \beta}A_\alpha \partial_\nu A_\beta\partial_{\mu} \theta \right]
\nonumber
\\
&-&\int d^4x \tilde{A}_\mu(x) J^\mu(x) ,
\end{eqnarray}
where we integrated by parts the third term in the r.h.s. of (\ref{EA-2}).  $J^\mu(x)=(J^0,\mathbf{J})$ represents the contribution of the ordinary (non-anomalous) electric four-current, determined by the one-loop tadpole diagrams. 

The Euler-Lagrange equations derived from the action (\ref{EA}) give rise to the modified Maxwell equations 
\begin{eqnarray} %\label{axionMaxwell}
&\mathbf{\nabla} \cdot \mathbf{E}=J^0+\frac{e^2}{4\pi^2}qB, \label{1}
 \\
&\nabla \times \mathbf{B}-\partial \mathbf{E}/\partial t=\mathbf{J}-\frac{e^2}{4\pi^2} \mathbf{q}\times \mathbf{E},  \label{2}
\\
&\mathbf{\nabla} \cdot \mathbf{B}=0, \quad \nabla \times \mathbf{E}+\partial \mathbf{B}/\partial t=0 \label{3},
\end{eqnarray} 
on which we already used that $\theta=qz/2$. These are the equations of axion electrodynamics for the MDCDW phase, which are a particular case of those proposed by Wilczek \cite{axionElect} many years ago for a general axion field $\theta$. 

It can be seen from equations (\ref{1})-(\ref{2}) that the axion term leads to an anomalous electric charge density, 
\begin{equation}\label{anomalous-Charge}
J_{anom}^0=\frac{e^2}{4\pi^2}qB,
\end{equation}
as well as to an anomalous Hall current density, 
\begin{equation}\label{anomalous-Current}
\mathbf{J}_{anom}=-\frac{e^2}{4\pi^2} \mathbf{q}\times \mathbf{E}
\end{equation}
The anomalous Hall current is perpendicular to both, the magnetic and the electric field, since q is aligned with $\mathbf{B}$. Besides, $\mathbf{J}_{anom}$ is dissipationless and as such, it can significantly influence the transport properties of the system. 

\section{Topological and LLL Origin of the Anomalous Electric Four-Current}

\subsection{Connection to the Chiral Anomaly}
The topological origin of the anomalous electric four-current can be clearly seen from the fact that the term that produces the anomalous electric four-current is also responsible for the chiral anomaly, which is topological in nature as explained below. To see this, let us go back to the expression (\ref{Z-theta}) and consider $\partial_\mu \theta$ as a background axial gauge field. We can derive the equation for the axial current by taking the functional derivative of the action with respect to the axial gauge field, and then take the divergence of the result to show that the axial four-current $J_5^{\mu}$ obeys the anomalous continuity equation

\begin{equation}\label{chiralanomaly}
\partial_\mu J^\mu_5=\frac{\kappa}{8} F_{\mu\nu}\tilde{F}^{\mu \nu}
\end{equation}

The presence of a mass just add a term to this equation, but the anomalous contribution will still be present. Going to Euclidean space does not change the form of this equation. Now, notice that the r.h.s. of (\ref{chiralanomaly}) can be written as the total derivative of a Chern-Simons current
\begin{equation}\label{cher-simonsJ}
\frac{\kappa}{8} F_{\mu\nu}\tilde{F}^{\mu \nu}=\partial_{\mu}K^{\mu},
\end{equation}
with
\begin{equation}\label{cher-simonsJ}
K^{\mu}=-\frac{\kappa}{16}\epsilon^{\mu\nu\rho\lambda} A_\nu \partial_\rho A_{\lambda},
\end{equation}
so if we integrate (\ref{chiralanomaly}) in the Euclidean four-volume, and use (\ref{cher-simonsJ}), the four-dimensional integral of $\frac{\kappa}{8} F_{\mu\nu}\tilde{F}^{\mu \nu}$ can be written as the integral of the Chern-Simons current $K^{\mu}$ over the surface of the four-dimensional volume. Such an integral gives an integer known as the winding number or topological charge of the gauge field.

\begin{equation}\label{winding-numb}
n\equiv \frac{\kappa}{8}\int d^4x F_{\mu\nu}\tilde{F}^{\mu \nu}
\end{equation}

From the regularization procedure discussed in Section 3 we have 

\begin{eqnarray}\label{index1}
\frac{\kappa}{4}\int d^4x F_{\mu\nu}\tilde{F}^{\mu \nu}&=&\lim_{M \rightarrow \infty} \int d_E^4x \mathrm{tr}\sum_n\phi_n^{\dag}(x) i \gamma_5 e^{-\slashed{D}^\dag(\mu,\theta)\slashed{D}(\mu,\theta)/M^2}\phi_n(x) \nonumber
\\
&+&\lim_{M \rightarrow \infty} \int d_E^4x\mathrm{tr}\sum_n \widetilde{\phi}_n^{\dag}(x) i \gamma_5 e^{-\slashed{D}(\mu,\theta)\slashed{D}^\dag(\mu,\theta)/M^2}\widetilde{\phi}_n(x)\nonumber
\\
&=&2n
\end{eqnarray}

Now, taking into account that $\gamma_5$ commutes with $\slashed{D}^\dag(\mu,\theta)\slashed{D}(\mu,\theta)$ and with $\slashed{D}(\mu,\theta)\slashed{D}^\dag(\mu,\theta)$, the eigenfunctions of these operators have defined chirality. If $\gamma_5\phi_n=\pm\phi_n$ then $\gamma_5\widetilde{\phi}_n=\gamma_5\lambda^{-1}_n \slashed{D}(\mu,\theta)\phi_n=-\lambda^{-1}_n \slashed{D}(\mu,\theta)\gamma_5\phi_n=\mp\widetilde{\phi}_n$, so for nonzero $\lambda_n$, the eigenfunctions $\phi_n$ and $\widetilde{\phi}_n$ always have opposite chirality. This implies that all the contributions of the nonzero singular modes cancel out in (\ref{index1}). 

For the zero modes, following similar derivations to those performed in Section 3, one finds
\begin{equation}\label{Pontryagin}
\int d^4x \frac{\kappa}{8} F_{\mu\nu}\tilde{F}^{\mu \nu}=n= \frac{1}{2}[\textrm{index}(i \slashed{D}(\mu,\theta)) +\textrm{index}(i \slashed{D}^\dag(\mu,\theta))]
\end{equation}
where $\textrm{index}(i \slashed{D}(\mu,\theta))=n_R-n_L$, $\textrm{index}(i \slashed{D}^\dag(\mu,\theta))=\widetilde{n}_L-\widetilde{n}_R$ are the indexes of these operators, defined as the difference between the number of zero modes with right (left) and left (right) chirality.  Here we used that $\textrm{index}(i \slashed{D}(\mu,\theta)) =\textrm{index}(i \slashed{D}^\dag(\mu,\theta))$.

The above results establish the topological nature of the axion term in the action and its connection to the chiral anomaly. Given that the anomalous electric four-current is obtained from this same term, it is also topological. Eq. (\ref{Pontryagin}) is an example of the Atiyah-Singer index theorem \cite{Nakahara}. As pointed out by several authors \cite{FujikawaPRD21_1980,Zyuzin-Burkov}, the quantity $\frac{\kappa}{8}F_{\mu\nu}\widetilde{F}^{\mu\nu}$ can then be interpreted as an index "density." 

\subsection{LLL Origin of the Anomalous Contributions}

We are now ready to discuss an independent way to establish the topological nature of the anomalous quantities in Eqs. (\ref{1})-(\ref{2}), and their connection to the asymmetry of the LLL spectrum. For that, we need to consider the Atiyah-Patodi-Singer invariant $\eta$ \cite{AS}, a topological quantity that measures the amount of spectral asymmetry of a theory. This invariant is defined as $\eta=\sum_k \mathrm{sgn}(E_k)$, where $E_k$ are the energy modes of the single-particle Hamiltonian \cite{NS} of the problem. 

We can readily observe two important things about $\eta$. First, the sum in the modes is ill-defined and needs proper regularization. Second, only the modes asymmetric about zero can contribute to $\eta$. In the case under study here, the energies $E_k$ are given by Eqs. (\ref{LLLspectrum})-(\ref{HighLspectrum}), and since the only asymmetric modes are the LLL ones, $\eta$ will be purely due to the LLL spectrum.

Using the methods discussed in \cite{NS} for topological theories of fermions at finite temperature and/or density, the regularized Atiyah-Patodi-Singer index can be found from\begin{equation}\label{etaindex}
\eta=\lim_{s\to0}\sum_k \mathrm{sgn}(E_k)|E_k|^{-s}.  
\end{equation}

As explicitly shown in \cite{PLB743}, the asymmetry of the LLL modes in the MDCDW phase yields $\eta\neq 0$ and as a consequence, the quark number acquires a topological (anomalous) contribution 
\begin{equation}\label{Quark-Number-Topology}
N^{anom}=-\eta/2=\int d^3x \rho^{anom}=\int d^3x\sum_{f}\rho^{anom}_f=\int d^3x \frac{3|e|}{4\pi^2}Bq
\end{equation}
\begin{equation}\label{Quark-Number-Flavor}
\rho^{anom}_f=N_c\frac{|e_f|}{4\pi^2}Bq, 
\end{equation}
with $\rho_f^{anom}$ denoting the anomalous quark number density of each flavor and $N_c$ the color number. In the next section, we will see that the same topological contribution can be found employing a different regularization procedure, first introduce in Ref. \cite{KlimenkoPRD82}, that allows to extract the anomalous part of the thermodynamic potential and then use it to find the anomalous quark number density. The extension of this calculation to the isospin asymmetric case was done in \cite{PRD92}.

We know that the LLL spectrum is effectively one-dimensional (i.e. independent of the transverse momentum). Since only the LLL contributes to the index $\eta$, we  could think that the nontrivial topology is a consequence of the dimensional reduction of the LLL modes. After all, the regularizations followed in \cite{KlimenkoPRD82} and \cite{PLB743} work because, after the limit of the regulator is taken ($\Lambda\to \infty$ in \cite{KlimenkoPRD82}, $s \to 0$ in \cite{PLB743}), the result is finite and independent of the regularization parameter, but such regulator-independence is attained precisely because the integral in momentum is only one-dimensional. However, the dimensional reduction of the LLL is always present in a background magnetic field, but in the absence of the DCDW condensate the LLL modes are symmetric, so in this case they cannot produce a nonzero $\eta$. Is the topology then rather a consequence of the inhomogeneous condensate? The answer is no too. The spectrum of the DCDW phase at zero magnetic field is symmetric \cite{DCDW}, so $\eta$ vanishes and no anomalous quark number exists in this case either. Only when the DCDW condensate is formed in the presence of a magnetic field, the LLL modes become asymmetric and their contribution leads to the nonzero $\eta$. Therefore, the nontrivial topology emerges in the MDCDW phase as the interplay of the DCDW ground state \emph{and} the dimensional reduction of the LLL modes. 

How does this translate into the anomalous electric charge? If we now multiply the topological quark number density of each flavor (\ref{Quark-Number-Flavor}) by its electric charge and then sum in flavor, we obtain the same anomalous electric charge density 
\begin{equation}\label{charge-density}
J^0_{anom}= \sum_fe_f \rho^{anom}_f=3\frac{e_u^2-e_d^2}{4\pi^2} Bq=\frac{e^2}{4\pi^2} Bq,
 \end{equation} 
that is found directly from the axion term in the electromagnetic action: the anomalous contribution to (\ref{1}). This result underlines not just the topological origin of the anomalous electric charge, but as importantly, its connection to the spectral asymmetry of the LLL energies.

Since the anomalous electric charge and the anomalous Hall current come from the same chiral anomaly term in the action, it is safe to say that they both originate from the spectral asymmetry of the LLL modes.

\section{Ordinary Charges and Currents}

We just established the LLL origin of the anomalous electric four-current density $J_{anom}=(J^0_{anom}, \mathbf{J}_{anom})$. Could the  anomalous four-current be cancelled out by the ordinary one? This important question is going to be explored in this section. Given that the anomalous four-current gets contributions only from the asymmetric modes (LLL modes), we shall only need to be concerned with the LLL contribution to the ordinary four-current. In addition, as previously stated, the anomalous terms are of one-loop order (as should be since they come from a fermion determinant \cite{Bilal}). Therefore, in the following derivations, we will find the LLL contribution to the one-loop ordinary four-current. Obviously, in principle there are also higher Landau level contributions to the ordinary four-current since we are not assuming here a particular strength of the magnetic field, but those contributions are not relevant for the goal of this section.
 
\subsection{LLL Quark Propagator in the MDCDW Phase}

The LLL contributions to the ordinary electric charge and currents can be found from the one-loop tadpole diagrams with internal lines of LLL fermions. We will also see that the ordinary Hall current can be readily found once we have the ordinary charge density. 

We first need to obtain the LLL propagator of each quark flavor. From (\ref{U_1-MF_L}), we can easily extract the LLL inverse quark propagator in the background magnetic field
\begin{equation}\label{inverse-propagator}
G^{-1}_{LLL}(k)=\gamma_\|^\mu(\tilde{k}_\mu^\|+\mathrm{sgn}\left(e_f\right)\frac{q}{2} \delta_{\mu 3}\gamma^5)-m,
\end{equation}
where $\tilde{k}_\mu^\|=(k_0-\mu, k_3)$, and $\mathrm{sgn}\left(e_f\right)$ distinguishes the contributions from $u$ and $d$ quarks. The LLL propagator $G_{LLL}(k)$ must satisfy 
\begin{equation}\label{product-propagators}
G^{-1}_{LLL}(k)G_{LLL}(k)=G_{LLL}(k)G^{-1}_{LLL}(k)=I,
\end{equation}
and it can be proposed as
\begin{equation}\label{propagator-1}
G_{LLL}(k)=\frac{AB}{\det G^{-1}_{LLL}(k)}
\end{equation}
with
\begin{eqnarray}\label{A}
A&=&-\gamma^5G^{-1}_{LLL}(k)\gamma^5\nonumber
\\
&=&\gamma_\|^\mu(\tilde{k}_\mu^\|+\mathrm{sgn}\left(e_f\right)\frac{q}{2} \delta_{\mu 3}\gamma^5)+m
\end{eqnarray}

After multiplying 
\begin{equation}\label{A-2}
G^{-1}_{LLL}(k)\cdot A=(\tilde{k}_\|^2+\frac{q^2}{4}-m^2)I+2\mathrm{sgn}\left(e_f\right)\frac{q}{2} \tilde{k}_0(i\gamma^1\gamma^2),
\end{equation}
it is easy to check that 
\begin{equation}\label{B}
B=(\tilde{k}_\|^2+\frac{q^2}{4}-m^2)I-2\mathrm{sgn}\left(e_f\right)\frac{q}{2} \tilde{k}_0(i\gamma^1\gamma^2),
\end{equation}
substituted in (\ref{propagator-1}) makes $G_{LLL}(k)$ to satisfy (\ref{product-propagators}).

Then, substituting with (\ref{A}) and (\ref{B}) in (\ref{propagator-1}) we obtain
\begin{equation}\label{propagator-3}
G_{LLL}(k)=\frac{\gamma_\|^\mu \tilde{k}_\mu^++m}{(\tilde{k}_0^{+})^2-\varepsilon^2}\Delta(+)+\frac{\gamma_\|^\mu \tilde{k}_\mu^-+m}{(\tilde{k}_0^{-})^2-\varepsilon^2}\Delta(-),
\end{equation}
where $\tilde{k}_\mu^{\pm}=(k_0-\mu\pm\mathrm{sgn}\left(e_f\right)\frac{q}{2}, 0, 0, k_3)$, $\gamma_\mu^\|=(\gamma_0,0,0,\gamma_3)$ and $\varepsilon=\sqrt{k_3^2+m^2}$ and the spin projectors are given by $\Delta(\pm)=(I\pm i\gamma^1\gamma^2)/2$.

Keeping in mind that the quarks in the LLL have only one spin projection (parallel/antiparallel to the field for positive/negative charged quarks), and that we have taken $\mathbf{B}$ in the positive $z$ direction, we can write the LLL propagator of flavor $f$ as
\begin{equation}\label{propagator}
G^f_{LLL}(k)=G_{LLL}(k)\Delta(\mathrm{sgn}\left(e_f\right)),
\end{equation}
with the spin projector denoted by $\Delta(\mathrm{sgn}\left(e_f\right))=(1+\mathrm{sgn}\left(e_f\right) i\gamma^1\gamma^2)/2$.

\subsection{LLL Electric Charge and Currents}
At finite temperature, the tadpole diagram contributes to the LLL ordinary four-current density of each flavor as
\begin{equation}\label{4-current}
J^\mu_{LLL}(\mathrm{sgn}\left(e_f\right))=(ie_f)\frac{|e_fB|N_cT}{(2\pi)^2}\sum_{n=-\infty}^{\infty} \int_{-\infty}^{\infty} dk_3 tr \left [i\gamma^\mu G^E_{LLL}(k) \right],
\end{equation}
 where all the quantities must be understood to be in the Euclidean space, so that the index $\mu=(1,2,3,4)$, $\gamma^\mu=\gamma_\mu$ are the Euclidean gamma matrices, $G^E_{LLL}(k)$ is the propagator (\ref{propagator-3}) after changing to Euclidean variables, and $k_4=\frac{(2n+1)\pi}{\beta}$.

Taking the trace we have 
\begin{equation}\label{trace-0}
tr \gamma_{4} [\gamma_4(k_4+i(\mu-q/2))+\gamma_3k_3-m]\Delta(\pm) =-2[k_4+i(\mu-q/2)],
\end{equation}
\begin{equation}\label{trace-1}
tr \gamma_{1,2}[\gamma_4(k_4+i(\mu-q/2))+\gamma_3k_3-m]\Delta(\pm)=0,
\end{equation}
\begin{equation}\label{trace-2}
tr\gamma_3[\gamma_4(k_4+i(\mu-q/2))+\gamma_3k_3-m]\Delta(\pm)=-2k_3.
\end{equation}

It is clear that the LLL does not contribute to the ordinary electric current density, since $J_{LLL}^{1,2}=0$, due to the zero trace (\ref{trace-1}), and $J_{LLL}^3$ is also zero after integrating in $k_3$. 

On the other hand, the LLL ordinary Euclidean electric charge density is 
\begin{equation}\label{4-current-LLL}
J^4_{LLL}(\mathrm{sgn}\left(e_f\right))=\frac{e_f|e_fB|N_cT}{2\pi^2}\sum_{k_4}  \int_{-\infty}^{\infty} dk_3 \frac{k_4+i(\mu-q/2)}{[k_4+i(\mu-q/2)]^2+\varepsilon^2}. 
\end{equation}

We can now carry out the Matsubara sum in (\ref{4-current-LLL}) and use $J^4=-iJ^0$ to find the LLL ordinary electric charge density
\begin{eqnarray}\label{4-current-T}
J^0_{LLL}(\mathrm{sgn}\left(e_f\right))=\frac{-e_f|e_fB|N_c}{2\pi^2} \int_{-\infty}^{\infty} dk_3 & [  n_F(\varepsilon+\mu-q/2)
\nonumber
\\
 &-n_F(\varepsilon-\mu + q/2) ],
\end{eqnarray}
where $n_F(x)=[1+\exp(\beta x)]^{-1}$ is the Fermi-Dirac distribution.

Taking the zero-T limit, integrating, and summing in flavor, we find that the LLL contribution to the ordinary electric charge density in the medium is
\begin{eqnarray}\label{regular-charge}
J_{LLL}^0&=&\sum_f J^0_{LLL}(\mathrm{sgn}\left(e_f\right))
\\
&=&\frac{e^2B}{2\pi^2}\sqrt{(\mu-q/2)^2-m^2}[\Theta(\mu-q/2-m)-\Theta(q/2-\mu-m)]\nonumber
\end{eqnarray} 

Notice that this result is not the same as the anomalous electric charge density (\ref{anomalous-Charge}), hence it does not cancel the anomalous contribution in the Maxwell equation. Only if we would put $m=0$ in (\ref{regular-charge}), meaning setting the condensate amplitude to zero, the anomalous electric charge density will be cancelled by (\ref{regular-charge}) in Eq. (\ref{1}). In such a situation, the resulting LLL contribution to the net electric charge density reduces to $\frac{e^2B}{2\pi^2}\mu$, a non-anomalous term which, as expected, is independent of $q$ since no physical quantity should depend on $q$ when there is no MDCDW condensate.

Finally, since the Maxwell equation (\ref{2}) contains an anomalous Hall current, it is important to investigate if it can be cancelled out by an  ordinary Hall current. However, before we tackle this problem, it is convenient to establish a simple formula to extract the Hall conductivity from the electric charge density, when the last one is linear in the magnetic field. 

Consider an electric charge density linear in the magnetic field $J^0=\sigma B_z$, with B pointing along $z$ and $\sigma$ some function of the condensate parameters and the chemical potential but independent of the electromagnetic field. Such an electric charge contributes to the effective action with a term $\int dt dV A_0J^0$. We then can write
\begin{eqnarray}\label{HC-derivation}
\int dt dV A_0J^0&=&\int dt dV A_0\sigma B_z=\int dt dV A_0\sigma (\partial_y A_x- \partial_x A_y)\nonumber
\\
&=&\int dt dV (E_y\sigma A_x-E_x\sigma A_y)\nonumber
\\
&=&\int dt dV (J^{Hall}_xA_x+ J^{Hall}_yA_y)
\end{eqnarray}
where $J^{Hall}_x=\sigma E_y $, $J^{Hall}_y=-\sigma E_x$ are Hall currents in the $x$ and $y$ direction respectively with Hall conductivity $\sigma_{xy}\equiv \sigma$, and we assumed static fields. 

One can then obtain the Hall conductivity as
\begin{equation}\label{HC-formula}
\sigma_{xy}=\frac{\partial J^0}{\partial B},
\end{equation}
which is known as the Str\v{e}da formula \cite{Streda}. Applying this formula to the electric charge density (\ref{regular-charge}), the LLL contribution to the ordinary Hall conductivity is
\begin{equation}\label{ordHC}
\sigma^{ord}_{xy}= \frac{\partial J_{LLL}^0}{\partial B} =\frac{e^2}{2\pi^2}\sqrt{(\mu-q/2)^2-m^2}[\Theta(\mu-q/2-m)-\Theta(q/2-\mu-m)]
\end{equation}
which leads to the LLL ordinary Hall current $\mathbf{J}_{LLL}^{ord}=(\sigma^{ord}_{xy}E_y,-\sigma^{ord}_{xy}E_x,0)$.  Clearly, $\mathbf{J}_{LLL}^{ord}$ does not cancel out the anomalous current (\ref{anomalous-Current}).

Likewise, the anomalous Hall conductivity can be found either from the anomalous charge
\begin{equation}\label{anomHC}
\sigma^{anom}_{xy}= \frac{\partial J_{anom}^0}{\partial B} =\frac{e^2}{4\pi^2}q,
\end{equation} 
or directly from the anomalous Hall current $\mathbf{J}_{anom}$ given in (\ref{anomalous-Current}). As $J_{anom}^0$ is due to the LLL, so is $\sigma^{anom}_{xy}$, thereby underlining once again the LLL origin of  $\mathbf{J}_{anom}$.
\subsection{LLL Quark Number Density}

In this subsection, we would like to find the ordinary quark number density directly from the thermodynamic potential, and also use it to obtain the ordinary electric charge density, so to check the result (\ref{regular-charge}) with an independent method.

With this goal, we start from the thermodynamic potential of the MDCDW phase in the one-loop approximation 
\begin{equation}\label{Thermodynamic-Pot}
\Omega=\Omega_{vac}(B)+\Omega_{anom}(B,\mu)+\Omega_\mu(B,\mu)+\Omega_T(B, \mu, T)+\frac{m^2}{4G}, 
\end{equation}
Here $\Omega_{vac}$ is the vacuum contribution; $\Omega_{anom}$ is the anomalous contribution, extracted from the LLL part of the medium term after proper regularization  \cite{KlimenkoPRD82}; $\Omega_\mu$ is the zero-temperature medium contribution and $\Omega_T$  the thermal part. For a single quark flavor they are 

\begin{eqnarray}\label{TP-Contributions}
\Omega^f_{vac}&=&\frac{1}{4\sqrt{\pi}} \frac{N_c|e_fB|}{(2\pi)^2}\int_{-\infty}^\infty dk\sum_{l\xi\epsilon} \int^{\infty}_{1/\Lambda^2} \frac{ds}{s^{3/2}}e^{-s(E)^2}
\\
\Omega^f_{anom} &=&- \frac{N_c|e_fB|}{(2\pi)^2} q\mu
\\
\Omega^f_\mu&=&-\frac{1}{2} \frac{N_c|e_fB|}{(2\pi)^2}\int_{-\infty}^\infty dk\sum_{\xi,l>0}2\lbrack(\mu-E)\Theta(\mu-E)\rbrack\vert_{\epsilon=+}\nonumber
\\
&\quad&+\Omega^{fLLL}_\mu
\\
\Omega^f_T&=& -\frac{1}{2} \frac{N_c|e_fB|}{(2\pi)^2}\int_{-\infty}^{\infty} dk\sum_{l\xi\epsilon}\ln \left (1+e^{-\beta (|E-\mu|} \right )
\end{eqnarray} 
with $E$ the energy modes (\ref{LLLspectrum}) and (\ref{HighLspectrum}), and
\begin{eqnarray}\label{Medium-Contribution-2}
\Omega^{fLLL}_\mu&=&-\frac{1}{2} \frac{N_c|e_fB|}{(2\pi)^2}\int_{-\infty}^{\infty} dk\sum_\epsilon(|E^0-\mu|-|E^0|)_{reg}
\\
&=&-\frac{N_c|e_fB|}{(2\pi)^2} \Bigg\{\Bigg[ Q(\mu) + m^2\ln \bigg(m/R(\mu)\bigg)\Bigg ] \Theta(q/2-\mu-m)\Theta(q/2-m)\nonumber
\\
&-&\Bigg[ Q(0) + m^2\ln \bigg(m/R(0)\bigg)\Bigg ] \Theta(q/2-m)\nonumber
\\
&+&\Bigg[ Q(\mu) + m^2\ln \bigg(m/R(\mu)\bigg)\Bigg ]  \Theta(\mu-q/2-m)\nonumber
\\
&-&\Bigg[ Q(0) + m^2\ln \bigg(m/R(0)\bigg)\Bigg ]  \Theta(\mu-q/2-m)\Theta(-q/2-m)\Bigg\},\nonumber
\end{eqnarray}
the LLL contribution to the medium part. Here, we introduced the notation
\begin{eqnarray}\label{Parameter definitions}
Q(\mu)&=&|q/2-\mu|\sqrt{(q/2-\mu)^2-m^2}, \quad \quad Q(0)=|q/2|\sqrt{(q/2)^2-m^2}\nonumber
\\
R(\mu)&=&|q/2-\mu|+\sqrt{(q/2-\mu)^2-m^2}, \quad R(0)=|q/2|+\sqrt{(q/2)^2-m^2}\nonumber
\end{eqnarray}

Notice that $\Omega^f_{anom}$ favors a nonzero modulation $q$, as it decreases the free-energy of the system. We recall that such a term exists thanks to the asymmetry of the LLL modes. 

The quark number density can be found from the derivative of $\Omega$ with respect to the baryon chemical potential. At $T=0$, the quark number density of each flavor has two contributions, one anomalous
\begin{equation}\label{Anomalous-Charge-density}
\rho_f^{anom}=-\frac{\partial \Omega^f_{anom}}{\partial \mu}= \frac{N_c|e_fB|}{(2\pi)^2} q,
\end{equation}
which arises only from the LLL and coincides with (\ref{Quark-Number-Flavor}), and one ordinary, obtained from
\begin{equation}\label{Ordinary-Charge-density-2}
\rho_f^{ord}=-\frac{\partial \Omega^f_{\mu}}{\partial \mu}
\end{equation}

Since we are interested in the LLL contribution to the ordinary quark number density, we can use (\ref{Medium-Contribution-2}) to show that
\begin{eqnarray}\label{Ordinary-Charge-density-f}
\rho_{fLLL}^{ord}&=&-\frac{\partial \Omega^{fLLL}_\mu}{\partial \mu}
\\
&=&- \frac{N_c|e_fB|}{(2\pi)^2} \left [2\sqrt{(q/2-\mu)^2-m^2}[\Theta(q/2-\mu-m)-\Theta(\mu-q/2-m)] \right ]\nonumber
\end{eqnarray}

After summing in flavor we find
\begin{equation}\label{Ordinary-Charge-density}
\rho_{LLL}^{ord}=\frac{3|eB|}{2\pi^2}\sqrt{(\mu-q/2)^2-m^2}[\Theta(\mu-q/2-m)-\Theta(q/2-\mu-m)],
\end{equation}

Just as we did for the anomalous charge, we can similarly obtain the ordinary LLL electric charge from
\begin{eqnarray}\label{Ordinary-Quark-Charge-density}
J^0_{LLL}&=&\sum_{e_f}e_f\rho_{fLLL}^{ord}
\\
&=&\frac{e^2B}{2\pi^2}\sqrt{(\mu-q/2)^2-m^2}[\theta(\mu-q/2-m)-\theta(q/2-\mu-m)],\nonumber
\end{eqnarray}
As expected, this result reproduces the expression (\ref{regular-charge}) that was calculated directly from the tadpole diagrams.

\section{Magnetoelectricity in the MDCDW Phase}

The MDCDW phase exhibits linear magnetoelectricity. This can be seen by defining the $\mathbf{D}$ and $\mathbf{H}$ fields as 
\begin{equation}\label{def}
\mathbf{D}=\mathbf{E}-\kappa\theta \mathbf{B}, \quad \mathbf{H}=\mathbf{B}+\kappa\theta \mathbf{E}
\end{equation}
and then rewriting the Maxwell equations (\ref{1}) and (\ref{2}) in terms of the fields in the MDCDW medium,
\begin{equation}\label{eqs}
\mathbf{\nabla} \cdot \mathbf{D}=J^0,\quad \nabla \times \mathbf{H}-\frac{\partial \mathbf{D}}{\partial t}=\mathbf{J} 
\end{equation} 

Physically this means that a magnetic field induces an electric polarization $\mathbf{P}=-\kappa\theta\mathbf{B}$ and an electric field induces a magnetization $ \mathbf{M}=-\kappa\theta\mathbf{E}$. This is possible because, as seen from (\ref{def})-(\ref{eqs}), the MDCDW ground state breaks P and T-reversal symmetries. The magnetoelectricity here is different from the one found in the magnetic-CFL  phase of color superconductivity \cite{MCFL}, where P was not broken and the effect was a consequence of an anisotropic electric susceptibility \cite{ME-MCFL}, so it was not linear. 

The anomalous Hall current can then be found from an effective, medium-induced, magnetic current density $\nabla \times \mathbf{M}$ due to the space-dependent anomalous magnetization coming from the axion term.

\section{Anomalous transport in the MDCDW phase}
 
The MDCDW phase exhibits quite interesting properties. The most important is the existence of the dissipationless anomalous Hall current (\ref{anomalous-Current}) perpendicular to $\mathbf{E}$ and to the  modulation vector $\mathbf{q}$, which in turn is parallel to $\mathbf{B}$. We already proved that the anomalous Hall conductivity is given by
\begin{equation}\label{Hall-conductivity}
\sigma^{anom}_{xy}=e^2q/4\pi^2
\end{equation} 
Its anomalous character is reflected in the fact that it does not depend on the fermion mass $m$, consistent with the nondissipative character of the anomalous Hall current.

The same expression of the anomalous Hall conductivity has been found in WSM \cite{Zyuzin-Burkov}, where the role of the modulation parameter $q$ is played by the separation in momentum of the Weyl nodes. A similar Hall conductivity can appear also at the boundary between a topological and a normal insulator \cite{Qie-PRB78} when there is an electric field in the plane of the boundary. However, in the topological insulator case, the anomalous Hall conductivity is discrete because the axion field $\theta$ jumps from $0$ to $\pi$ in the surface of the two insulators.  Our results are also connected to optical lattices, as 3D topological insulators have been proposed to exist in 3D optical lattices \cite{PRL105}.

It is worth to point out the relevance of these results for neutron stars \cite{UniverseEF-VI}. If the quark density in neutron stars  is high enough to accommodate MDCDW matter threaded by a poloidal magnetic field, then, any electric field present in the medium, whether due to the anomalous electric charge or not, and as long as it is not parallel to the magnetic field, will lead to dissipationless Hall currents in the plane perpendicular to the magnetic field. 
The existence of this kind of current could serve to resolve the issue with the stability of the magnetic field strength in magnetars  \cite{Bstability}. In another direction, it will be important to understand if this new magneto-transport property can significantly affect the thermal and electric conductivity producing a tangible separation between the transport properties of compact stars formed by neutrons or by quarks in the MDCDW phase. 
We underline that the condition of electrical neutrality does not need to be satisfied locally for compact hybrid stars \cite{Glendenning}, which could have a core in the MDCDW phase with an anomalous charge contribution and Hall currents circulating inside and at the surface.
These and other questions highlight the importance to explore which observable signatures could be identified and then used them as telltales of the presence of the MDCDW phase in the core. 

The anomalous Hall current could be also produced in future HIC like those planned at the Nuclotron-based Ion Collider Facility (NICA) at Dubna, Russia \cite{PRC85} and at the Facility for Antiproton and Ion Research (FAIR) at Darmstadt, Germany \cite{1607.01487}, which will explore the high density, cold region of the QCD phase map, and where event-by-event off-central collisions will likely generate perpendicular electric and magnetic fields \cite{NICA}.  It will be interesting to carry out a detailed quantitative analysis of how these currents could lead to observable signatures, even after taking into account that there the QGP distributes itself more as an ellipsoid than as an sphere about the center of the collision. The Hall currents will tend to deviate the quarks from the natural outward direction from the collision center and one would expect a different geometry of the particle flow in the MDCDW phase compared to other dense phases that have no anomalous electric current. The realization of the MDCDW phase in the QGP of future HIC experiments is likely viable because the inhomogeneity of the phase is characterized by a length $\Delta x = \hbar / q \sim 0.6 fm$ for $q \sim \mu =300$ MeV \cite{KlimenkoPRD82}, much smaller than the characteristic scale $L\sim 10 fm$ of the QGP at RHIC, NICA,  and FAIR, while the time scale for this phase will be the same as for the QGP. 

Other interesting effects might emerge by considering the fluctuations $\delta\theta$ of the axion field. If one goes beyond the mean-field approximation, there will be mass and kinetic terms of the axion field fluctuation.  Besides, due to the background magnetic field, the axion fluctuation couples linearly to the electric field via the term $\kappa\delta\theta \mathbf{E}\cdot \mathbf{B}$, so the field equations of the axion fluctuation and the electromagnetic field will be mixed, giving rise to a quasiparticle mode known as the axion polariton mode \cite{axpolariton}. The axion polariton mode is gapped with a gap proportional to the background magnetic field. This implies that electromagnetic waves of certain frequencies will be attenuated by the MDCDW matter, since in this medium they propagate as polaritons. The axion polariton could be useful to design a way to probe the presence of the MDCDW phase in future HIC experiments at high baryon densities, due to its effect in the attenuation of certain light frequencies when light is shined through the collision region.

\section{Concluding Remarks}

In this paper we studied the topological effects of the MDCDW phase, which could be one of the phases of cold-dense quark matter in a magnetic field. We showed that the system exhibits an anomalous charge that depends on the applied magnetic field and the modulation of the particle-hole condensate. The topological nature of the electric charge can be traced back to the spectral asymmetry of the LLL modes. The spectral asymmetry is also responsible for an anomalous non-dissipative Hall current that depends on the modulation parameter.

We call the reader's attention to an interesting connection between ultraviolet (UV) and infrared (IR) phenomena in the MDCDW phase. The appearance of $\Omega_{anom}$ in the thermodynamic potential (\ref{TP-Contributions}) is a consequence of the regularization of the high-energy modes in the difference of two ill-defined sums from which the anomalous and the finite medium contributions are extracted \cite{KlimenkoPRD82}. Since the anomalous term contributes to the gap equation for $q$, whose origin is IR because it comes from the quark-hole pairing, we have that the UV physics affects the IR properties of the system.

The results we are reporting can have significance for HIC physics and neutron stars. Future HIC experiments, that will take place at lower temperatures and higher densities, will certainly generate strong magnetic and electric fields in their off-central collisions and will open a much more sensitive window to look into a very challenging region of QCD. For example, the Compressed Baryonic Matter (CBM) at FAIR \cite{1607.01487}  have been designed to run at unprecedented interaction rates to provide high-precision measures of observables in the high baryon density region. That is why it is so timing and relevant to carry out detailed theoretical investigations of all potential observables of the MDCDW phase. Therefore, we hope that our findings will serve to stimulate quantitative studies to identify signatures of the anomalous effects here discussed in the future HIC experiments. 

Interestingly, the anomalous effects of the MDCDW phase share many properties with similar phenomena in condensed matter systems with non-trivial topologies as topological insulators \cite{Qie-PRB78}, where $\theta$ depends on the band structure of the insulator; Dirac semimetals \cite{Dirac-SM}, a 3D bulk analogue of graphene with non-trivial topological structures; and WSM \cite{Weyl-SM}, where the derivative of the angle $\theta$ is related to the momentum separation between the Weyl nodes. Countertop experiments with these materials can therefore help us to gain useful insight of the physics governing the challenging region of strongly coupled QCD, thereby inspiring new strategies to probe the presence of the MDCDW phase in neutron stars and HIC. 

An important question that should be tackled in detail in the near future is the stability of the MDCDW phase. Can the MDCDW phase be erased by the fluctuations of the condensate at arbitrary small temperatures, as known to occur in the real kink crystal phase \cite{Instability} or in the DCDW phase \cite{Instability-2}? We anticipate that the answer is no. Let us explain why. In the DCDW phase, the energy spectrum of the Nambu-Goldstone fields has soft-energy modes in the transverse momenta that produce infrared divergencies and hence erase the long-range order of the condensate at any finite temperature \cite{Instability-2}. The problematic fluctuations are of two types. One is a mix of the phonon and the chiral fluctuation in the third internal direction. The corresponding Goldstone boson is electrically neutral. The other consists of  chiral fluctuations in the internal directions 1 and 2. These fluctuations are electrically charged. For the neutral fluctuation we expect that, just as was argued in the appendix of \cite{Instability}, the external magnetic field should qualitatively modify the dispersion of the phonon/chiral mix in such a way that it will be linear in all the directions, thus eliminating the infrared divergences. On the other hand, for the charged fluctuations the modification of the dispersion should be due to the fact that they couple with the background magnetic field through the covariant derivative. Similarly to what we proved several years ago for the charged fluctuations in the Magnetic CFL phase \cite{MCFLphases}, the dispersions of the charged fluctuations in the MDCDW phase must acquire a field-dependent mass because in the presence of the magnetic field the chiral symmetry is explicitly reduced to the subgroup $U(1)_L\times U(1)_R $, so there are no charged Goldstone bosons.

\textbf{Acknowledgments:}
This work was supported in part by NSF grant PHY-1714183, and PSC-CUNY Award  60650-00 48.


\begin{thebibliography}{00}

\bibitem{QCDreviews} K.  Fukushima and T. Hatsuda, Rept. Prog. Phys. {\bf74} (2011) 014001;  M. Buballa, Phys.  Rep. \textbf{407} (2005) 205.

\bibitem{Aoki06} Y. Aoki et al., Phys. Lett. B \textbf{643} (2006) 46; JHEP \textbf{0906} (2009) 088; H. Abuki \textit{et. al}, Phys. Rev. D, \textbf{78} (2008) 034034.

\bibitem{NJL} Y. Nambu and G. Jona-Lasinio, Phys. Rev. \textbf{122}
(1961)  345; ibid. \textbf{124} (1961) 246.

\bibitem{largeNQCD} D. V. Deryagin, D. Y. Grigoriev and V.  A. Rubakov, Int. J. Mod. Phys. A \textbf{7 (}1992) 659; E. Shuster and D. T. Son, Nucl. Phys. B \textbf{573} (2000) 434; B.-Y. Park et. al, Phys. Rev. D \textbf{ 62} (2000) 034015.

\bibitem{q-chiralspirals}T. Kojo, Y. Hidaka, L. McLerran and R. D. Pisarski, Nucl. Phys. A \textbf{843} (2010) 37; T. Kojo,
et al., Nucl. Phys. A \textbf{875} (2012) 94; T. Kojo, R.D. Pisarski and A. M. Tsvelik, Phys. Rev. D \textbf{82} (2010) 074015; T. Kojo, Nucl. Phys. A \textbf{877} (2012) 70.

\bibitem{NickelPRD80} D. Nickel, Phys. Rev. D \textbf{80} (2009) 074025; Phys. Rev. Lett. \textbf{103} (2009) 072301.

\bibitem{largeN}R. Rapp, E. Shuryak, and I. Zahed, Phys. Rev. D \textbf{63} (2001) 034008; N. V. Gubina \textit{et. al}, Phys. Rev. D \textbf{86} (2012) 085011.

\bibitem{PRD82-054009} S. Carignano, D. Nickel, and M. Buballa, Phys. Rev. D \textbf{82} (2010) 054009.

\bibitem{PRD85-074002} H. Abuki, D. Ishibashi, and K. Suzuki, Phys. Rev. D \textbf{85} (2012) 074002.

\bibitem{Klevansky92} S. Klevansky, Rev. Mod. Phys. \textbf{64} (1992) 649.

\bibitem{CS-MRP80} M. Alford, A. Schmitt and K. Ragagopal, Rev. Mod. Phys. \textbf{80} (2008) 1455.

\bibitem{CS-MRP86} R. Anglani, R. Casalbuoni, M. Ciminale, N. Ippolito, R. Gatto, M. Mannarelli and M. Ruggieri, Rev. Mod. Phys. \textbf{86} (2014) 509.

\bibitem{gCFL} R. Casalbuoni, R. Gatto, M. Mannarelli, G. Nardulli and M. Ruggieri, Phys. Lett. B \textbf{605} (2005) 362; K. Fukushima, Phys. Rev. D \textbf{72} (2005) 074002.

\bibitem{g2SC} M. Huang, and I. A. Shovkovy, Phys. Rev. D \textbf{70} (2004) 094030; I. Giannakis and H.-C. Ren, Phys. Lett. B \textbf{611} (2005) 137.

\bibitem{TBinCS} S. Reddy and G. Rupak, Phys. Rev. C \textbf{71} (2005) 025201; K. Fukushima, Phys. Rev. D \textbf{73} (2006) 094016; M. Hashimoto, Phys. Lett. B \textbf{642} (2006) 93; M. Huang, Phys. Rev. D \textbf{73} (2006) 045007.

\bibitem{LO} A. I. Larkin and Y. N. Ovchinnikov, Sov. Phys. JETP  \textbf{20} (1965) 762.

\bibitem{FF} P. Fulde and R. A. Ferrell, Phys. Rev. \textbf{135} (1964) A550.


\bibitem{CSLOFF} M. G. Alford, J. A. Bowers and K. Rajagopal, Phys.Rev. D \textbf{63} (2001) 074016; J. A. Bowers and K. Rajagopal, Phys. Rev. D \textbf{66} (2001) 065002; R. Casalbuoni and G. Nardulli, Rev. Mod. Phys. \textbf{76} (2004) 263.

\bibitem{gluonCS} E. J. Ferrer and V. de la Incera, Phys. Rev. D \textbf{76} (2007) 114012.


\bibitem{Ferrer-Incera-PLB} E. J. Ferrer and V. de la Incera, Phys. Lett. B \textbf{769} (2017) 208.

\bibitem{Odyniec}G. Odyniec, J. Phys.: Conf. Ser. \textbf{455} (2013) 012037.


\bibitem{1607.01487}  T. Ablyazimov et. al. arXiv:1607.01487.


\bibitem{PRC85}W-T. Deng and X-G. Huang, Phys. Rev. C \textbf{85} (2012) 044907; V. Toneev, O. Rogachevsky and V. Voronyuk, Eur. Phys. J. A \textbf{52} (2016) 264.

\bibitem{Nuclear-matter-field} L. Dong and S. L. Shapiro ApJ. \textbf{383} (1991) 745.


\bibitem{Quark-Matter-field} E. J. Ferrer, V. de la Incera, J. P. Keith, I. Portillo and P. L. Springsteen, Phys. Rev. C \textbf{82} (2010) 065802.

\bibitem{CME} D. E. Kharzeev, L. D. McLerran and H. J. Warringa, Nucl. Phys. A\textbf{ 803} (2008) 227;
K. Fukushima, D. E. Kharzeev and H. J. Warringa, Phys. Rev. D \textbf{78} (2008) 074033.


\bibitem{Skokov}V. Skokov, A.Yu. Illarionov and V. D. Toneev, Int. J. Mod. Phys. A \textbf{24} (2009) 5925.

\bibitem{Wen} E. J. Ferrer, V. de la Incera and X-J Wen, Phys. Rev. D \textbf{91} (2015) 054006.





\bibitem{KlimenkoPRD82} I. E. Frolov, V. Ch. Zhukovsky and K. G. Klimenko, Phys. Rev. D \textbf{82} (2010) 076002.

\bibitem{PLB743} T. Tatsumi, K. Nishiyama and S. Karasawa, Phys. Lett. B \textbf{743} (2015) 66.

\bibitem{AMM} E. J. Ferrer, V. de la Incera, I. Portillo and M. Quiroz Phys. Rev. D \textbf{89} (2014) 085034. 

\bibitem{ferrer-incera} B. Feng, E. J. Ferrer and V. de la Incera, Nucl. Phys. B \textbf{853} (2011) 213; Phys. Lett. B  \textbf{706} (2011) 232; 

\bibitem{Lec-Notes}E. J. Ferrer and V. de la Incera, Lect. Notes Phys. \textbf{871} (2013) 399; arXiv:1208.5179 [nucl-th].

\bibitem{ferrer-incera-sanchez} E. J. Ferrer, V. de la Incera and A. Sanchez, Acta Phys. Polon. Supp. \textbf{5} (2012) 679.



\bibitem{Weyl-SM} A. A. Burkov and I. Balents, Phys. Rev. Lett. \textbf{107} (2011) 127205; A. A. Zyuzin and A. A. Burkov, Phys. Rev. B \textbf{86} (2012) 115133.


\bibitem{burkov-2017} A. A. Burkov, arXiv:1704.06660 [cond-mat.mess-hall].

\bibitem{Overhauser} A. W. Overhauser, Adv. Phys. \textbf{27} (1978) 343.

\bibitem{QqM}  L. McLerran and R. D. Pisarski, Nucl. Phys A \textbf{796} (2007) 83.

\bibitem{InhCRev} M. Buballa and S. Carignano, Prog. Part. Nucl. Phys. \textbf{81} (2015) 39.


\bibitem{DCDW}E. Nakano and T. Tatsumi, Phys. Rev. D \textbf{71} (2004) 114006.

\bibitem{FujikawaPRD21_1980} K. Fujikawa, Phys. Rev. Lett. \textbf{42} (1979) 1195; Phys. Rev. D \textbf{21} (1980) 2848.

\bibitem{Fujikawa_book} K. Fujikawa, Path Integrals and Quantum Anomalies. Clarendon Press, Oxford, 2004.

\bibitem{JHEP1112_2011}T. Kanazawa, T. Wettig, and N. Yamamoto, JHEP \textbf{1112} (2011) 007.

\bibitem{Nakahara} M. Nakahara, \textit{Geometry, Topology and Physics}, Graduate Student Series in Physics, Ed. D. F. Brewer, IOP Publishing Ltd 1990.

\bibitem{Bilal} A. Bilal, \textit{Lectures on Anomalies}, LPTENS-08-05, arXiv:0802.0634 [hep-th].

\bibitem{anomalywithmu} A. Gomez Nicola and R. Alvarez-Estrada, Int. J. Mod. Phys. A \textbf{9} (1994) 1423;
S. D. H. Hsu, F. Sannino, and M. Schwetz, Mod. Phys. Lett. A \textbf{16} (2001) 1871; R. V. Gavai and S.
Sharma, Phys. Rev. D \textbf{81} (2010) 034501.

\bibitem{Zyuzin-Burkov} A. A. Zyuzin and A. A. Burkov, Phys. Rev. B \textbf{86} (2012) 115133.


\bibitem{VI-LAquila} V. de la Incera, J. Phys. Conf. Ser. \textbf{861} (2017) 012019.


%\bibitem{Fuji-mu} A. Gomez Nicola and R. Alvarez-Estrada, Int. J. Mod. Phys. A \textbf{9} (1994) 1423; S. D. H. Hsu, F. Sannino, and M. Schwetz, Mod. Phys. Lett. A \textbf{16} (2001) 1871;  T. Kanazawa, T. Wettig, and N. Yamamoto, JHEP \textbf{1112} (2011) 007; S. Shi, W-M Sun, and H-S Zong,  Modern Physics Letters A, \textbf{28} (20013) 1350006.

%\bibitem{PRD81}R. V. Gavai and S. Sharma, Phys. Rev. D \textbf{81} (2010) 034501.


\bibitem{axionElect} F. Wilczek, Phys. Rev. Lett. \textbf{58} (1987) 1799.


%\bibitem{jackiw-Ano} R. Jackiw, Int. J. Mod. Phys. A  \textbf{25} (2010) 659.


\bibitem{AS} 
M. Atiyah, V. Patodi and I. Singer, Proc. Cambridge Phils. Soc. \textbf{77} (1975) 42; \textbf{78} (1995) 405; \textbf{79} (1976) 71.

\bibitem{NS}
A. J. Niemi, Nucl. Phys. B  \textbf{251} (1985) 155;
A. J. Niemi and G. W. Semenoff, Phys. Reports \textbf{135} (1986) 99.

\bibitem{PRD92} S. Carignano, E. J. Ferrer, V. de la Incera and L. Paulucci, Phys. Rev. D \textbf{92 }(2015) 105018.

\bibitem{Streda} P. Str\v{e}da 1982 J. Phys. C: Solid State Phys. \textbf{15}  L717.

\bibitem{MCFL} E. J. Ferrer, V. de la Incera and C. Manuel, Phys. Rev. Lett. \textbf{95} (2005) 152002; Nucl. Phys. B \textbf{747} (2006) 88; J. Phys. A  \textbf{39} (2006) 6349. 


\bibitem{ME-MCFL} B. Feng, E. J. Ferrer and V. de la Incera, Phys. Lett. B \textbf{706} (2011) 232.


 \bibitem{Qie-PRB78}X-L Qi, T. L. Hughes and S-C Zhang, Phys. Rev. B \textbf{78} (2008) 195424.

\bibitem{PRL105} A. Bermudez, L. Mazza, M. Rizzi, N. Goldman, M. Lewenstein and M. A. Martin-Delgado, Phys. Rev. Lett. \textbf{105} (2010) 190404.

\bibitem{UniverseEF-VI}E. J. Ferrer and V. de la Incera, Universe \textbf{4} (2018)  54.

\bibitem{Bstability}
A. K. Harding and D. Lai, Rept. Prog. Phys. \textbf{69} (2006) 2631; H. C. Spruit, AIP Conf. Proc. \textbf{983} (2008) 391.

\bibitem{Glendenning}
N. K. Glendenning, Compact Stars, Nuclear Physics, Particle Physics, and General Relativity, Springer-Verlag, NY 2000.


\bibitem{NICA}
E. J. Ferrer and V. de la Incera, Eur. Phys. J. A \textbf{52} (2016) 266.


\bibitem{axpolariton}
R. Li, J. Wang, X-L Qi and S-C Zhang, Nature Phys. \textbf{6}  (2010) 284.


\bibitem{Dirac-SM} S. M. Young, S. Zaheer, J. C. Y. Teo, C. L. Kane, E. J.
Mele and A. M. Rappe, Phys. Rev. Lett., \textbf{108} (2012) 140405; S. Borisenko, Q. Gibson, D. Evtushinsky, V. Zabolotnyy, B. Buchner and R. J. Cava, Phys. Rev. Lett. \textbf{113} (2014) 027603; M. Neupane, et al, Nature Communications, \textbf{5} (2014) 3786; Z. K. Liu, et al, Nature Mat. \textbf{13} (2014) 677.

\bibitem{Instability} Y. Hidaka, K. Kamikado, T. Kanazawa and T. Noumi, Phys. Rev. D \textbf{92} (2015) 034003.

\bibitem{Instability-2} T. G. Lee, E. Nakano, Y. Tsue, T. Tatsumi and B. Friman, Phys. Rev. D \textbf{92} (2015) 034024.
 
\bibitem{MCFLphases} E. J. Ferrer and V. de la Incera, Phys. Rev. D \textbf{76} (2007) 045011.




\end{thebibliography}
\end{document}